\documentstyle[prl,aps,epsf]{revtex}
\begin{document}
\draft
\twocolumn[\hsize\textwidth\columnwidth\hsize\csname @twocolumnfalse\endcsname

\title{
Commensurate and Incommensurate Vortex Lattice Melting in 
Periodic Pinning Arrays}  
\author{C.~Reichhardt} 
\address{Center for Nonlinear Studies and Applied Theoretical and 
Computational Physics Division, Los Alamos National Laboratory, Los Alamos, NM
87545} 

\author{C.J.~Olson}
\address{ Theoretical and Applied Physics Divisions, Los Alamos
National Laboratory, Los Alamos, NM 87545}

\author{R.T.~Scalettar and G.T.~Zim\'anyi} 
\address{Department of Physics, University of California---Davis, Davis, CA
95616} 

\date{\today}
\maketitle
\begin{abstract}
We examine the melting of commensurate and incommensurate vortex lattices
interacting with square pinning arrays through the use of 
numerical simulations. 
For weak pinning strength in the commensurate case
we observe an order-order 
transition from a commensurate square vortex lattice to a triangular
floating solid phase as a function of temperature. 
This floating solid phase 
melts into a liquid at still higher temperature. For strong
pinning there is only a single transition from the square pinned lattice
to the liquid state. 
For strong pinning in the incommensurate case, 
we observe a multi-stage melting in which the interstitial vortices
become mobile 
first,
followed by the melting of the entire lattice, 
consistent with recent imaging experiments. 
The initial motion
of vortices in the incommensurate phase occurs by an exchange process
of interstitial vortices with vortices located at the pinning sites.
We have also examined the vortex melting behavior for higher matching
fields and find that a coexistence of a commensurate pinned vortex
solid with an interstitial vortex liquid occurs while at higher 
temperatures the entire vortex lattice melts. 
For triangular arrays at incommensurate fields higher than the 
first matching field we observe that the initial vortex motion
can occur through a novel correlated ring excitation where 
a number of vortices
can rotate around a pinned vortex. 
We also discuss the relevance of our results to recent experiments
of colloidal particles interacting with periodic trap arrays. 
\end{abstract}

\vskip2pc]

\section{Introduction}

Vortices interacting with periodic pinning arrays have attracted
growing interest with the advances of nano-lithographic techniques in which
the size and shape of individual
pinning sites, as well as the geometry of the pinning array,
can be carefully controlled.
These pinning arrays can be composed of 
1D periodic modulations \cite{Daldini},
micro-holes 
\cite{Fiory,Metlushko,Meltlush,Harada,Moschalkov,Bending,Field} 
or magnetic dots \cite{Schuller,Fassano,Hoffman} 
in which the holes or dots have a diameter $d$ 
significantly less than the dot spacing $a$. The holes and dots  
suppress the order parameter, allowing the flux lines
in the mixed state to lower their energy, and thereby pinning the vortices. 
Strong commensurability effects in these systems are  
observed at the matching fields, when  the number of 
particles equals an integer multiple of the number of divots 
\cite{Fiory,Metlushko,Meltlush,Schuller,Reichhardt}. 
Transport measurements near 
$T_{c}$ above the first matching field indicate that some of the
vortices  are only weakly pinned in the interstitial regions. 
These vortices become mobile at a lower applied current
than the vortices at the pinning sites \cite{Moschalkov}.     
Direct evidence for commensurate states and interstitial vortices 
has been obtained using Lorentz force microscopy \cite{Harada} and
scanning Hall probe microscopy \cite{Bending,Field}. 
Magnetization measurements by Baert {\it et al.} \cite{Metlushko} 
interpreted the 
loss of pinning above the first matching field near $T_{c}$ as due
to a coexistence of an interstitial vortex liquid with a pinned vortex
solid at the pinning sites. 

In addition to the technological applications of superconductors
which require strong pinning, vortices interacting with periodic
pinning arrays are also an ideal realization of an
elastic lattice interacting with a
periodic substrate which can be generalized to a wide variety of
condensed
matter systems 
including atoms adsorbed on a surface \cite{Nelson}, colloidal crystals 
interacting with 
interfering laser beams\cite{Chowdhury}, and vortices in Josephson-junction
arrays \cite{Franz,Hattel}. 
A very similar system
to vortices in periodic arrays that has recently been
experimentally realized is colloidal particles
interacting with 
optical trap arrays \cite{Dufernese2,Dufrense,Korda,Frey}. 
Recent experiments
in these systems for colloids interacting with a square array observed
an interesting transition from a commensurate square colloidal crystal to a 
triangular colloidal crystal for increasing colloidal densities
when the colloidal interactions
began to dominate over the substrate \cite{Korda}. 

Although studies of the effects of temperature have been conducted on 
the dynamics of vortices interacting with periodic pinning arrays
\cite{Marconi}, the
static melting has not been investigated.
Here one could expect that 
vortex lattice melting behavior may be significantly different at
commensurate vs incommensurate fields.
Melting of vortices in wire networks has been studied \cite{Franz} for 
systems for filling fractions far less than one with Monte-Carlo
simulations.
The Monte-Carlo simulations of vortices in wire networks at 
low filling factors 
by Franz and Teitel \cite{Franz}
have shown a transition from a pinned solid 
to a floating solid phase, both of which have the same
symmetry as the underlying pinning lattice. 
For higher temperatures this floating solid
phase was found to melt into a liquid.   
In the wire network case there are no interstitial regions, 
and at the
matching field the floating solid phase is absent \cite{Franz}. 
The wire network system has several differences from vortices in
periodic
arrays in that in the former the vortices can be considered to move in
an
egg-carton potential. For vortices in periodic arrays the vortices
would
move in a
muffin-tin substrate in which there are flat potential regions between the
minima. In this system at low temperatures at commensurate matching,
particles in the overlying lattice will sit in the divots or energy  
minima, while at the incommensurate matching, 
some particles will sit in the {\it interstitial} regions 
between the 
divots \cite{Harada,Moschalkov,Reichhardt}. Even at the matching field
it can be possible for vortices to move into the interstitial regions
when the vortex-vortex interactions become dominant or thermal
fluctuations are strong enough.

Another difference between vortices in wire networks compared to vortices in
periodic pinning arrays appears for fields greater than the first
matching field. 
The additional vortices in the 
imaging experiments by Harada {\it et al.} \cite{Harada}
and simulations by Reichhardt {\it et al} \cite{Reichhardt}
indicate that in periodic pinning arrays
a wide variety of interstitial vortex crystals 
can be stabilized
at different matching fields.

The vortices located in the
interstitial regions can have significantly different static, dynamic
and thermal 
properties than the commensurate particles, and the coupling between the
two different species can give rise to interesting behavior.
Recent imaging experiments by Grigorenko {\it et al} \cite{Bending} 
using high resolution scanning
Hall probe microscopy near the matching field $B_{\phi}$ indicate that, 
for fields slightly 
below 
and above 
the first matching field,
over time the vacancies and interstitials appear to show site to site hopping.
The nature of the hopping dynamics is however not known, such as whether the
interstitial vortices jump from one interstitial site to another or if the
interstitials undergo an exchange process with the vortices at the pinning
sites. These experiments were done for square arrays and it is not known
how the vortex hopping might occur in different geometries such as
triangular pinning arrays. 

The melting of interstitial vortices which coexist with strongly
pinned vortices has also been proposed by Radzihovsky \cite{Leo} for
vortices interacting with randomly placed columnar defects where there
are more vortices than pinning sites.  



In this work we report the study of two-dimensional
melting of vortices interacting with a square muffin tin pinning potential
with the use of molecular dynamics simulations. 
For the commensurate case at low substrate strengths we observe a novel
order-order transition
from a square pinned vortex lattice with long-range order 
to a triangular ordered floating solid phase with quasi-long range order.  
Our simulations also allow us to examine the vortex dynamics at 
the transition into the floating solid where the vortices show a
collective 1D
avalanche motion along the pinning rows. For 
increased substrate strength the floating 
solid phase is lost and the pinned solid melts directly to a liquid.  
In the incommensurate case 
for vortex densities slightly above the first 
matching field for strong substrates we observe a 
multi-stage melting in which the initial vortex motion occurs by 
an exchange process of interstitial and pinned vortices. 

We have also examined the effects of temperature for higher matching
and non-matching fields greater than one. At the higher matching fields
we find that the vortices form a solid at low temperatures. For
higher temperatures we observe a coexistence of an interstitial vortex liquid
with a pinned vortex solid in agreement with the experiments of
Baert {\it et al}.  
At high incommensurate fields the extra vortices or vacancies in the
interstitial vortex liquid become mobile.
We observe the same general behaviors for triangular pinning arrays;
however, at the incommensurate fields the motion of extra vortices
or vacancies can occur by novel
collective ring excitations where a group of vortices rotate around
a vortex in a pinning site. These excitations occur when an
incommensurate number of vortices cage a vortex located at a pinning
site.

Our system should not only be of relevance for vortices and colloids
interacting with periodic substrates but also for the square to hexatic
vortex transitions observed in 
certain superconductors \cite{Thompson,Paul,Cubbit,Kogan}, where 
little is known
about the dynamics at the transitions. 

\section{Simulation}

We model a 2D system with periodic boundary conditions in $x$ and $y$
with $N_{v}$ vortices interacting with a square array of
$N_{p}$ pinning sites. We use a Langevin simulation for vortices
at finite temperature using the model of Brass and Jensen 
\cite{Jensen,Berlinsky}.
We consider the vortex motion to be overdamped and integrate the following
equation of motion.
\begin{equation}
{\bf f}_{i} = \eta{\bf v}_{i} = {\bf f}_{i}^{vv} + {\bf f}_{i}^{vp} + 
{\bf f}_{i}^{T}
\end{equation}
Here ${\bf f}_{i}$ is the total force acting on vortex $i$ and $\eta$ is the 
damping coefficient. The repulsive interaction 
of vortex $i$ with the other vortices is 
${\bf f}_{i}^{vv} = \sum_{j=1}^{N_{v}}f_{0}K_{1}(|{\bf r}_{i} - {\bf r}_{j}|/
\lambda){\bf {\hat r}}_{ij}$. Here,  
$\lambda$ is the penetration depth and 
$K_{1}(r/\lambda)$ is a modified 
Bessel function which falls off exponentially for $r > \lambda$, allowing 
us to place a cutoff in the  
interactions at $r = 6\lambda$ for computational
efficiency. The pinning is modeled as $N_p$ parabolic traps with 
${\bf f}_{i}^{vp} = -\sum_{k=1}^{N_p}(f_{p}/r_{p})(|{\bf r}_{i} - {\bf r}_{k}^{(p)} 
|)\Theta(r_{p} - |{\bf r}_{i} - {\bf r}_{k}^{(p)}|)
{\bf {\hat r}}^{(p)}_{ik}$.  
Here, $\Theta $ is the Heaviside step function, $r_{p}$ is the range of the
pinning potential, $f_{p}$ is the maximum pinning force of the wells,
${\bf {\hat r}}_{ij} = 
({\bf r}_{i} - {\bf r}_{j})/|{\bf r}_{i} - {\bf r}_{j}|$ and
${\bf {\hat r}}_{ik}^{(p)} = 
({\bf r}_{i} - {\bf r}_{k}^{(p)})/|{\bf r}_{i} - {\bf r}_{k}^{(p)}|$.
The thermal force has the properties $<f_{i}^{T}(t)> = 0$ and 
$<f_{i}^{T}(t)f_{j}^{T}(t^{'})> = 2\eta k_{B}T\delta_{ij}\delta (t - t^{'})$.
In this work we set $\eta = k_{B} = 1$. 
All lengths, fields and forces are given in units of 
$\lambda$, $\Phi_{0}/\lambda^{2}$ and $f_{0}$ respectively. 
For most of the results shown here the density of vortices is 
$0.25\Phi_{0}/\lambda^{2}$. The 
initial ground state of the system is 
found using simulated annealing.
We then
slowly increase $T$ and analyze the coordination number, $P_{n}$ 
(from the Voronoi or Wigner Seitz cell construction), and  
the 
structure factor, 
$$S({\bf k}) = (1/L^{2})\sum_{i,j}e^{i{\bf k}\cdot
[{\bf r}_{i} - {\bf r}_{j}]}$$ where $L$ is the system size length. 

Another measure we employ to characterize the thermally induced transition
is the sum of 
the displacements of the vortices from their initial positions at
$T = 0$ to their positions at a higher $T$:
$$d_{w}(T) = \sum_{i=1}^{N_{v}}|{\bf r}_{i}(T) - {\bf r}_{i}(0)|^{2}$$ 
If the vortex lattice configuration is the same as the initial state
than $d_{w} = 0$. If a portion of the vortices move but then become 
frozen again $d_{w}$ will jump up. If the vortex
lattice becomes molten $d_{w}$ will increase to a saturation value of
$L/2$.

To 
visualize vortex motion
we plot the trajectories that the vortices follow for a period of time. 
To examine the dynamic response in the phases we can add a driving term 
${\bf f}_{i}^{d}$ representing a Lorentz force from an applied current 
to Eq.~1 and sum over the net vortex velocities 
$V_{x} = \sum_{j=1}^{N_{v}}{\bf v}_{i}\cdot{\hat {\bf x}}$.  
To analyze finite size scaling in $S({\bf k})$ we conduct simulations  
with the same vortex and pin densities for
system sizes with a length $L$ for systems
ranging from  
$L = 24\lambda$ to $L = 60\lambda$ with $N_{v} = 144$ to $N_{v} = 900$. 

\section{Commensurate Melting}

In Fig.~1 we present the evolution of the coordination numbers $P_{6}$ and
$P_{4}$, the average power of the secondary peaks in the 
structure factor $<S({\bf k})>$, and the vortex 
displacements $d_{w}$ for increasing $T$ for a system at the matching field 
with $L = 48\lambda$. Here $f_{p} = 0.03f_{0}$, $r_{p} = 0.17\lambda$ and 
$a = 2.0\lambda$. 
For $ 0 < T < 0.0012$ the pinned vortex lattice has the same square
symmetry as the pinning lattice, as indicated by $ P_{4} \approx 1.0$. 
The initial increase in $d_{w}$ for $T< 0.0012$ is due to the vortices
moving randomly within pinning sites but remaining confined 
to a distance $r_{p}$.   
Near 
$T = 0.0012$ there is a structural transition in the vortex lattice
from square lattice to triangular, indicated by the sudden drop in 
$P_{4}$ to $\sim 0$ and the large increase in $P_{6}$ to $0.98$. 
In Fig.~1(b) we plot the average $<S({\bf k})>$ of the 
appropriate secondary peaks in
$S({\bf k})$. For $T < 0.0012$  we observe only four secondary peaks in
$S({\bf k})$ as indicated in the left inset in Fig.~1(a). 
At the square-triangular transition 
there is initially a large drop in $<S({\bf k})>$ from $0.95$ to $0.2$ and
then a recovery to $0.55$. The position of the secondary peaks shifts 
at the transition from four-fold to six-fold order. During the transition 
a mixture of {\it both} four and six-fold peaks appear,
reducing the overall power at each peak. In the triangular phase there is 
almost no power for the four-fold order. 
The square to triangular transition can also be observed 
as a jump in $d_{w}$ (Fig.~1(c)),
coinciding with the drops in $P_{4}$ and $<S({\bf k})>$, which is due to
the shifting of the vortex positions during the transition.   
In the triangular lattice phase, $d_{w}$ remains roughly constant,
indicating that diffusion is 
not occurring and that the vortex lattice is still a solid. Near 
$T =0.0049$ the triangular vortex lattice melts into a liquid,   
indicated by the increase in $d_{w}$ as well as the  drops in
$<S({\bf k})>$ and $P_{6}$. We note that $P_{6}$ will not drop to 
zero for a liquid state since even for random particle distributions
a portion of the vortices will always have six-fold coordination numbers.
We have also conducted simulations where we cool down in temperature starting
at $T = 0.008$. In this case we do not observe any hysteresis in the
liquid to triangular phase; however, hysteresis is present in the 
triangular to square transition, suggesting that the transition is first order
in nature. We have also examined the response of the phase to an applied drive
and find that only the square pinned phase has a finite critical depinning
force.   
  
The finite size scaling behavior 
of the structure factor 
$$
S({\bf k})/L^{2} \sim L^{-\theta} 
$$
which is described more fully in \cite{Franz}
gives us further information about the nature of the phases. 
For long range order, $\theta = 0$,
for floating solids or hexatic phases 
$0 < \theta < 2.0$, and for the liquid phase 
$\theta = 2.0$. 
For different system sizes
we average the power of the secondary peaks 
in $<S({\bf k})>$ for 150 frames. 
In the inset of Fig.~1(b) 
we plot $<S({\bf k})>/L^{2}$ (where $N_{v} \sim L^{2}$) versus 
$N_v$ for the 
three different phases.
In the commensurate phase ($T = 0.0005$) we find $\theta \approx 0.0$
corresponding to the pinned lattice. For $T = 0.003$, in the 
triangular phase, $\theta = 0.33 \pm 0.04$ which is consistent with a 
floating solid phase. In the floating solid phase we find a small 
variation of $\theta$ with $T$ which we will address 
elsewhere. In the liquid phase ($T = 0.007$) we find 
$\theta = 1.98 \pm 0.04 $. 

\subsection{Dynamics of the Order-Order Transition}

In order to gain insight into the dynamics of the square to triangular
transition, 
in Fig.~2  we show the real space images as well as the 
individual vortex trajectories in the pinned solid, the pinned-solid to 
floating solid transition, and the liquid phase for the system in Fig.~1. 
In the pinned solid phase shown in Fig.~2(a) ($T = 0.0005$) we find a 
square vortex lattice with little or no detectable vortex motion. 
Fig.~2(b) ($T = 0.00135$) shows the {\it dynamics} at the beginning of     
the pinned to floating solid transition. Here  
alternate rows of vortices   
move in a sudden 
1D avalanche where the vortices in one row move in a collective 
manner while the other rows remain
static. All the vortices in a given row shift together as the vortices
jump out of the pinning sites and move to the interstitial regions. 
The motionless rows 
shown in Fig.~2(b) also begin to move at 
a slightly later time. This shifting of the vortex positions  
produces the jump in $d_{w}$ 
at $T=0.0012$ seen in Fig.~1(c) at the square to floating
solid transition. 
In Fig.~2(c) ($T = 0.0015$), in the floating solid phase,  
the vortex lattice has triangular order and shows increased thermal 
wandering around the equilibrium positions as can be
seen in the vortex trajectories. This increased thermal wandering
also smears the secondary peaks in  $S({\bf k})$, reducing
the maximum value to $0.55$, as in Fig.~1(b).    
In Fig.~2(d) ($T = 0.0065$), 
in the molten state the vortices
are disordered and diffuse in a random manner 
corresponding to the large value of $d_{w}$ in Fig.~1(c). 

\subsection{Phase Diagram for the Commensurate Case}

By conducting a series of simulations with different substrate strengths
and using the criteria from Fig.~1 to determine the onset of the
different phases, we construct the phase diagram in Fig.~3. 
The range of temperatures over which the commensurate solid 
appears increases linearly for increasing $f_{p}$. 
The floating solid phase only appears for $f_{p} < 0.125f_{0}$. The 
line separating the floating solid to liquid phase is roughly constant
with temperature for lower substrate strengths at a value 
that corresponds to roughly the zero pinning melting temperature, 
$T_{m} = 0.0048$ (as also obtained with the diffusion and 
structure factor measurements).
All three phase boundaries can be understood from simple considerations.
The floating solid to liquid line is independent of $f_{p}$ 
because it is determined
by the vortex-vortex interactions and not the pinning force.
For $f_{p} \geq 0.125f_{0}$ the commensurate solid 
melts directly into a liquid and $<S({\bf k})>$ has the behavior shown 
in the inset of Fig.~3. The melting transition is seen to shift to 
{\it higher} temperatures with increasing pinning strength, 
as the melting occurs when  
thermal fluctuations are able to overcome the increasing pinning
strength. Finally the commensurate-to-floating solid follows 
$f_{p} \sim T$ with the low $f_{p}$ being washed out before the
vortex-vortex interactions are washed out. 
The nature of the floating solid to liquid transition, such 
as whether a dislocation-mediated melting transition occurs \cite{Nelson}, 
is beyond the scope of this work.

To further compare our results 
with those of vortices in wire networks \cite{Franz}, in which the
floating solid phase  
with the same symmetry as the pinned phase is observed as a function 
of commensurability, in our case the 
pinned to floating solid transition is seen as a
function of substrate strength. 
Further, we observe a square to triangular 
transition due to the fact that a portion of the vortices shift into  
pin free regions, which is not possible in wire networks. 

We note that in our simulation we imposed nearly 
square boundary conditions which
can create defects or stress in the triangular lattice.  
We do not
find any defects in  our triangular lattice and our finite size scaling is
consistent with that in \cite{Franz}. It is also possible that the 
square pinned solid phase may be enhanced for a higher temperature range 
by the square boundary conditions;
however, the general trends of the phase diagram should still hold.       

\section{Incommensurate Melting Near the First Matching Field}

We now consider the vortex behavior
above the first matching field when there are more vortices than pins. 
We find that for large substrate strengths, $f_{p} > 0.25f_{0}$,  
a commensurate sub-lattice forms, with the additional vortices 
sitting in the interstitial regions. 
In Fig.~4 we show the melting behavior for a system with 
$N_{v} = 1.063N_{p}$ and $f_{p} = 0.6f_{0}$.
In the imaging experiments of Grigorenko {\it et al.} \cite{Bending},
the interstitial motion was observed for a field of $B/B_{\phi} = 1.04$.
Here, for 
$T < 0.0016$ both the commensurate vortices and the interstitial vortices 
remain pinned as seen in Fig.~4(a). At these low temperatures, 
the interstitial vortices are pinned by the potential cage
created by vortices at the pinning sites. At $T > 0.0016$ vortex 
diffusion begins to occur as indicated by the increase
in $d_{w}$ in Fig.~5(b). This is due to the onset of motion of the    
incommensurations as seen in the vortex trajectories 
in Fig.4(b). 
The motion is {\it not} restricted to 
interstitial vortices. Instead, interstitial and pinned vortices 
{\it exchange places},
and over 
a period of time {\it all} the vortices take part in the motion. 
The exchange process can be understood by considering that the interstitial 
vortex will produce a force $f_{in}$ on the nearby commensurate
vortices which effectively reduces the pinning force to $|f_{p}| - |f_{in}|$. 
Since the melting transition is a function of the pinning force the 
melting temperature will be reduced.   

In Fig.~5(a) the onset of the incommensuration 
motion is also marked by a drop in $<S({\bf k})>$ from $0.96$ to 
approximately $0.7$. 
In contrast to the sharp drop in $<S({\bf k})>$ seen in the commensurate
case, here $<S({\bf k})>$ still retains strong four-fold order due to the
presence of the background lattice of pinned vortices as seen in Fig.~4(b). 
For higher temperatures $<S({\bf k})>$ gradually decreases 
with most order being lost for $T > 0.0175$, corresponding to the
melting temperature for the commensurate case with the same substrate strength. 
For simulations where we cool down from $T > 0.0175$ we do not observe any
hysteresis.
We have also conducted simulations in which we apply a constant 
drive of $f_{d} = 0.012$, gradually increase $T$ and 
measure the  
average vortex velocity $<V_{x}>$.    
For $T < 0.0016$, in the pinned interstitial phase $<V_{x}> = 0$. 
At the interstitial transition there is a jump in $<V_{x}>$ indicating
the onset of vortex motion.  
Finally at the overall vortex melting transition,  
$<V_{x}>$ jumps to a value corresponding to the entire lattice flowing. 

\section{Melting for Higher Commensurate and Incommensurate Fields}

\subsection{Commensurate Melting} 
We have also investigated the melting in square and triangular pinning
arrays for the higher matching fields 
$B/B_{\phi} = 2, 3, 4$, $5$ and $8$.
At these fields ordered vortex crystals are 
formed \cite{Harada,Reichhardt}. 
In Fig.~6 we show the vortex positions and trajectories for 
the square pinning array with $B/B_{\phi} = 4.0$ for increasing temperatures.
For $T < 0.004$ the vortex lattice is ordered as seen in experiments
and simulations. For $T>0.004$
[Fig.~6(b)] the interstitial vortex lattice melts; however, the vortices at the
pinning sites remain immobile so that a coexistence of a solid and a liquid
phase occurs. The interstitial vortices are constrained to move in a square
grid and there is a region around the pinning sites which the interstitial
vortices do not enter due to the repulsion of the vortices located at 
the pinning sites. Under an applied drive the depinning force is finite for
low temperatures but goes to zero at the onset of the interstitial vortex
liquid transition. This result is in agreement with the interpretations of the
experiments by Baert {\it et al.} in which the pinning force goes to zero
near $T_{c}$ when interstitial vortices are present, but is still finite
for fields at which vortices are only located at the pinning sites.  
For increased $T$, as in  Fig.~6(c), 
the vortices at the pinning sites become depinned
as well and the entire lattice is in the molten state.
The general behavior of the melting from Fig.~6 is also seen at the other
matching fields we have investigated with transitions from the solid, to 
solid-liquid coexistence, to the liquid state.

\subsection{Incommensurate Melting}
In Fig.~7 we show the melting behavior for the incommensurate field
$B/B_{\phi} = 4.1$. Here the onset of vortex diffusion occurs for a 
much lower temperature than for $B/B_{\phi} = 4.0$.
The vortex motion occurs at defect sites where additional interstitial
vortices are present
in the ordered $B/B_{\phi} = 4.0$ interstitial vortex crystal.
These extra interstitial vortices are less strongly pinned
than the other interstitial vortices. The vortex motion can occur
by the continuous motion of a single vortex over 
a certain distance, or by a pulse like motion in which a series of
individual vortices move by one lattice constant as
the pulse moves through.     
If the trajectories are drawn for a longer time period the
interstitial vortex
diffusion can be seen to occur through the entire lattice. 
As the temperature is increased the vortices at the pinning sites will
become depinned. 

\subsection{Collective Ring Excitations for Incommensurate Fields with
Triangular Pinning}

We have found that the melting behavior for 
square and triangular pinning arrays is similar
at the matching fields; 
however, at the incommensurate fields we observe clear differences
between the two pinning geometries.
In Fig.~8 we show the melting behavior for a system with triangular
pinning for $B/B_{\phi} = 3.08$. At low temperatures the system is frozen
but as $T$ is increased motion begins to occur in the form of ring 
excitations.   Here a portion of the vortices rotate around a
vortex at a pinning site; however, no net diffusion of
interstitial vortices is occurring. 
For $B/B_{\phi} = 3.0$ each vortex at a pinning
site is surrounded by six interstitial vortices. For fields above or below
$B/B_{\phi} = 3.0$ a portion of the pinned
vortices will be surrounded by 5 or 7 
vortices, and it is at these sites where the ring excitations occur. 
Due to the symmetry of the 
pinning array the vortices at the pinning sites create a 
potential around each pinning site which has six minima, so that at
$B/B_{\phi} = 3.0$ the vortex configuration is particularly stable. 
When there are seven or five vortices around 
a pinning sites, they are
incommensurate with the six mimima, and therefore 
the thermal kicks can cause the interstitial vortices to slowly rotate. 

As the applied field is moved further away from the matching field,
the number of incommensurations increases
and the number of ring excitations increases as well.
Such ring excitations are observed in triangular pinning arrays
for most incommensurate fields but with different numbers of vortices
forming the ring.  As the temperature increases, 
the initial diffusion of the vortices occurs via a hopping mechanism
in which the extra vortex in the rotating ring 
hops out of the ring and starts the rotation of a different ring. The
rate of this hopping increases with the temperature. 

\section{Summary}

In summary, we have investigated the melting of vortex lattices interacting 
with periodic pinning arrays for both the commensurate and incommensurate 
case. For the first matching field with weak substrates we have observed  
a transition in the vortex lattice from a 
square commensurate solid to a 
triangular floating solid which melts into a liquid at higher temperatures. 
This transition occurs by 
1D collective motion of vortices shifting 
from the pinning sites into the interstitial regions. 
For strong substrates the floating solid 
phase is lost and the vortex lattice melts directly into a liquid. 
For fields slightly above the first matching field  
for strong disorder we observe a multi-stage 
melting where the interstitial vortices are highly mobile consistent  
with recent imaging experiments. 
The motion of the interstitial vortices occurs 
by the exchange of 
interstitial with pinned vortices rather
than with the interstitials hopping from one 
interstitial site to another. As the temperature increases 
more vortices become mobile.

 We have also examined the melting behavior for matching fields greater
than one. At low temperature the interstitial vortices form a crystalline 
state.
At higher temperatures we observe a coexistence of an interstitial vortex liquid
with a commensurate pinned vortex lattice. The interstitial vortices can
diffuse in the regions between the pinned vortices. We also observe a 
depletetion zone around the pinned vortices which interstitial vortices do
not enter due to the repulsion from the pinned vortex. 
At higher temperatures the entire vortex lattice melts. At 
higher incommensurate fields the additional defects (extra interstitials
or vacancies) in the interstitial vortex crystal become mobile and diffuse
in the interstitial regions.      
With triangular pinning arrays at incommensurate fields the initial 
vortex motion occurs through a novel collective ring excitation where 
interstitial vortices can rotate around a pinned vortex. These rotations
occur when an incommensurate number of interstitial vortices surround a
pinned vortex. 

We briefly discuss some experimental systems in which these effects
could be observed.
For vortex lattices in periodic pinning arrays, vortex lattice melting is
most relevant to high Tc superconductors. Our results are only for a 2D system
whereas 3D effects can be relevant 
to melting. Our model would thus best describe 
high-temperature superconductors with periodic columnar defects where the
vortices would have line like behavior. 
However, several of the results here should still
be relevant for low temperature superconductors. For the low temperature 
superconductors melting would only occur very near $T_{c}$. For strong pinning
such as arrays of holes, a floating solid phase
at the first matching field would not be observable and 
the system should go directly from the pinned solid to the normal phase with
a very small region of a disordered vortex lattice. For weak pinning such as
small magnetic dots or weak defect arrays the pinned solid to floating solid
transition should occur well below $T_{c}$ and should be observable for
the low temperature superconductors. 
For a low temperature superconductor the phase diagram in Fig.~3 would
be modified, with the floating
solid to liquid transition replaced by the floating solid to
normal transition, although again a small vortex liquid state may occur 
just below $T_{c}$. Also the pinned solid to vortex liquid transition would
remain flat rather than increasing with $f_{p}$. 
The solid to floating solid transition could be imaged with 
scanning Hall probe measurements, neutron scattering, 
or Lorentz microscopy. In addition transport measurements would also be
able to reveal the loss of pinning at the transition.
     
The vortex behavior at the incommensurate fields should also be visible
in the low-temperature superconductors. The recent imaging experiments 
of Grigorenko {\it et al.} \cite{Bending} 
have already found evidence for the motion of
the highly mobile interstitial vortices in these types of systems.  
These same imaging techniques should
be able to image the pinned vortex lattice coexisting with the interstitial
vortex liquid. 
In triangular pinning arrays collective ring excitations would 
appear as a smeared ring around pinning sites with the Hall probe arrays
and the dynamics could be directly imaged with Lorentz microscopy.

Our work is also directly relevant to colloidal particles 
interacting with periodic pinning. 
A particularly promising realization of this system is optical trap
arrays in which the pinning strength 
can be easily tuned.  In such a system our phase diagram can be directly
tested.  
Recent experiments have already seen evidence for a square to triangular      
transition as the colloid density is increased
\cite{Korda}. In addition the dynamics
of the interstitial colloids can be directly imaged with video microscopy to
determine if the interstitials hop directly from 
one interstitial site to another, or if they move through an
exchange process with a pinned colloid. In addition it should be also be
possible to observe the melting behavior for higher matching colloidal
densities and the collective ring excitations for 
colloids interacting with triangular optical trap arrays. 

We thank S.~Bending, L.~DeLong, S.~Field, D.~Grier, N.~Gr{\o}nbech-Jensen, 
P.~Korda, V.~Moshchalkov, and G.~Spalding for useful discussions.  
Funding provided by NSF-DMR-9985978, CLC, CULAR, and
DOE grant W-7405-EBG-36.

\begin{figure}
\centerline{  
\epsfxsize=3.5in
\epsfbox{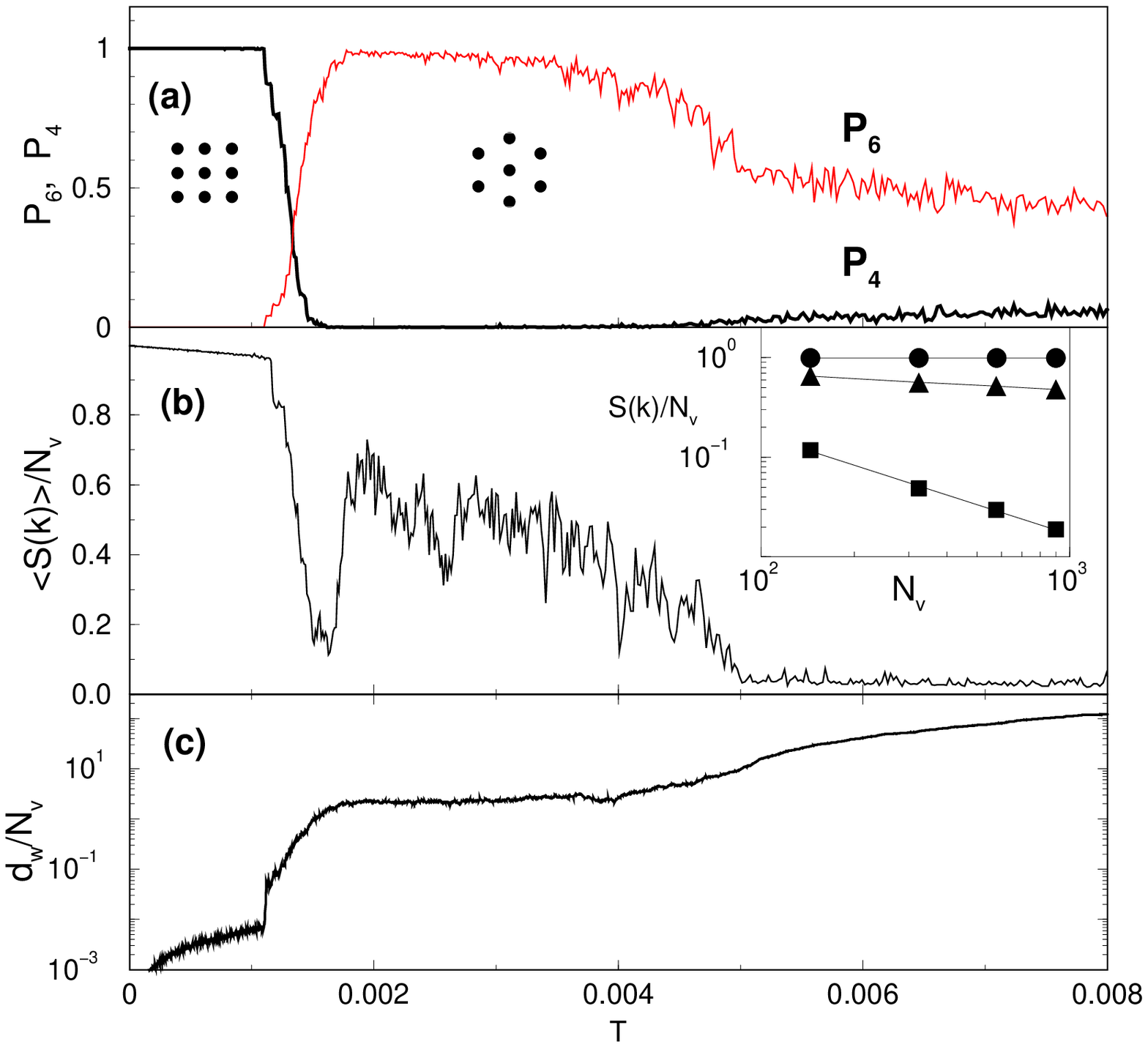}} 
\caption{
(a) The  fraction of four-fold $P_{4}$ (thick line) 
and six-fold $P_{6}$ coordinated (thin line) vortices 
versus $T$. (b) The average power of the secondary peaks, 
$<S({\bf k})>/N_{v}$ versus $T$.
The position of the peaks shifts from four-fold to six-fold near $T = 0.001$.
The peaks are lost at $T = 0.005$. The inset of (b) shows the scaling of the 
secondary peaks for different system sizes where $N_{v} \sim L^{2}$. For the
pinned phase (circles), $T = 0.0005$ and  $S(k) \sim L^{0}$; 
for the triangular lattice (triangles), $T = 0.003$ and 
$S(k) \sim L^{-0.33}$; and for the liquid phase (squares), 
$T = 0.007$ and $S(k) \sim L^{-1.98}$. (c) The displacements 
$d_{w}/N_{v} = <|{\bf r}(T) - {\bf r}(0)|^{2}>$ 
of the vortices versus $T$.}
\label{fig1}
\end{figure} 

\begin{figure}
\centerline{
\epsfxsize=3.5in
\epsfbox{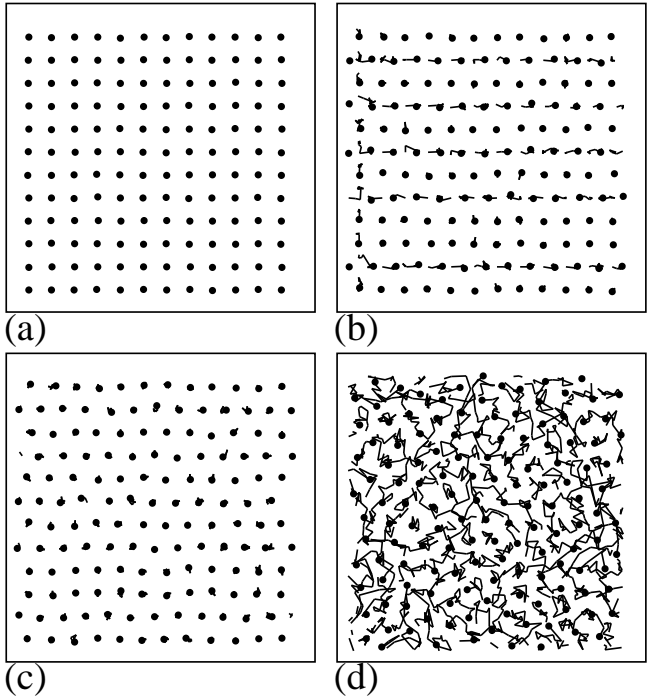}} 
\caption{
The vortex positions and trajectories for
a $24\lambda \times 24\lambda$ section of a $48\lambda\times48\lambda$ system 
for (a) the pinned solid ($T = 0.0005$),
(b) the beginning of the pinned solid to floating solid transition 
($T = 0.00135$), (c) the floating sold phase ($T = 0.0015$),
and (d) the liquid phase ($T = 0.0065$).  In (b) 
alternate rows of 
vortices shift positions. For slightly higher temperatures the 
other vortices also shift in position. This shifting is reflected in 
the jump in $d_{w}$ at $T=0.0012$ in Fig.~1(c).} 
\label{fig2}
\end{figure}

\begin{figure}
\centerline{
\epsfxsize=3.5in
\epsfbox{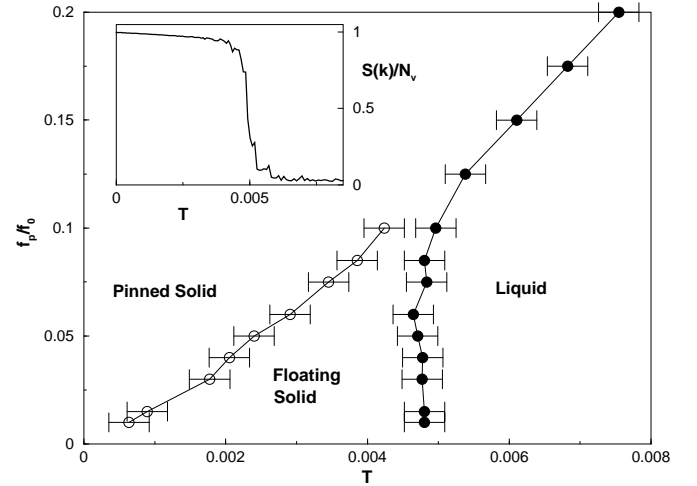}}
\caption{
The phase diagram of $f_{p}$ versus $T$. The open circles represent  
the pinned solid to floating solid transition. The filled circles 
represent the transition into the liquid phase. The inset shows the
behavior of $<S({\bf k})>/N_{v}$ versus $T$ for $f_{p} = 0.125f_{0}$ where 
the pinned solid melts directly to a liquid.} 
\label{fig3}
\end{figure}

\begin{figure}
\centerline{
\epsfxsize=3.5in
\epsfbox{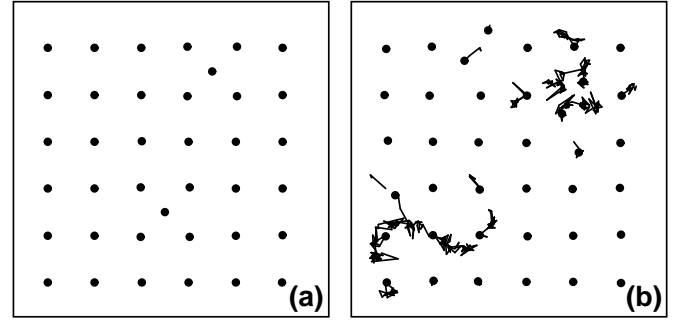}}
\caption{
(a) A snapshot of a $10\lambda \times 10\lambda$ region of the 
sample with $B/B_{\phi} = 1.06$.
The vortex positions and trajectories for $T = 0.0005$ show that some 
vortices are pinned at interstitial positions. (b) At $T = 0.0036$ the motion
of incommensurations can be seen while the background lattice remains 
pinned. The vortex trajectories in (b) show that vortex motion occurs by
the interstitial vortices pushing other vortices off the pinning
sites which in turn move into interstitial regions and repeat the process.}
\label{fig4}
\end{figure} 

\begin{figure}
\centerline{
\epsfxsize=3.5in
\epsfbox{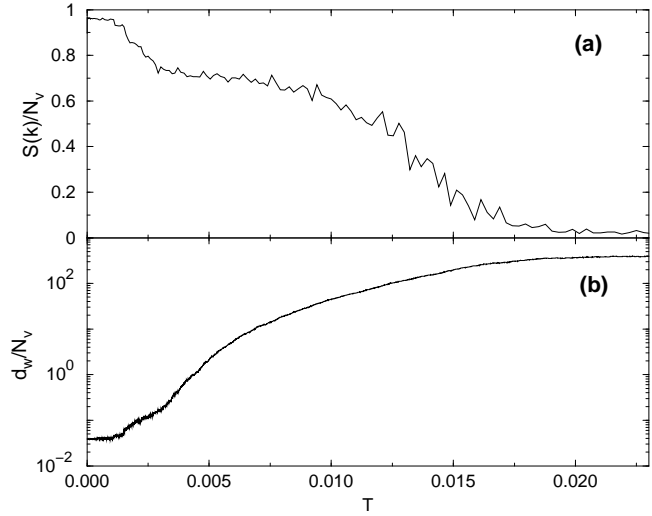}}
\caption{ 
The behavior of $<S({\bf k})>/N_{v}$ and $d_{w}/N_{v}$ 
versus $T$ for a system with
$f_{p} = 0.6f_{0}$ and $N_{v} = 1.063N_{p}$. Motion in the vortex lattice
starts at $T = 0.0016$ as indicated in the increase in $d$ (b). 
} 
\label{fig5}
\end{figure}

\begin{figure}
\centerline{
\epsfxsize=2.8in
\epsfbox{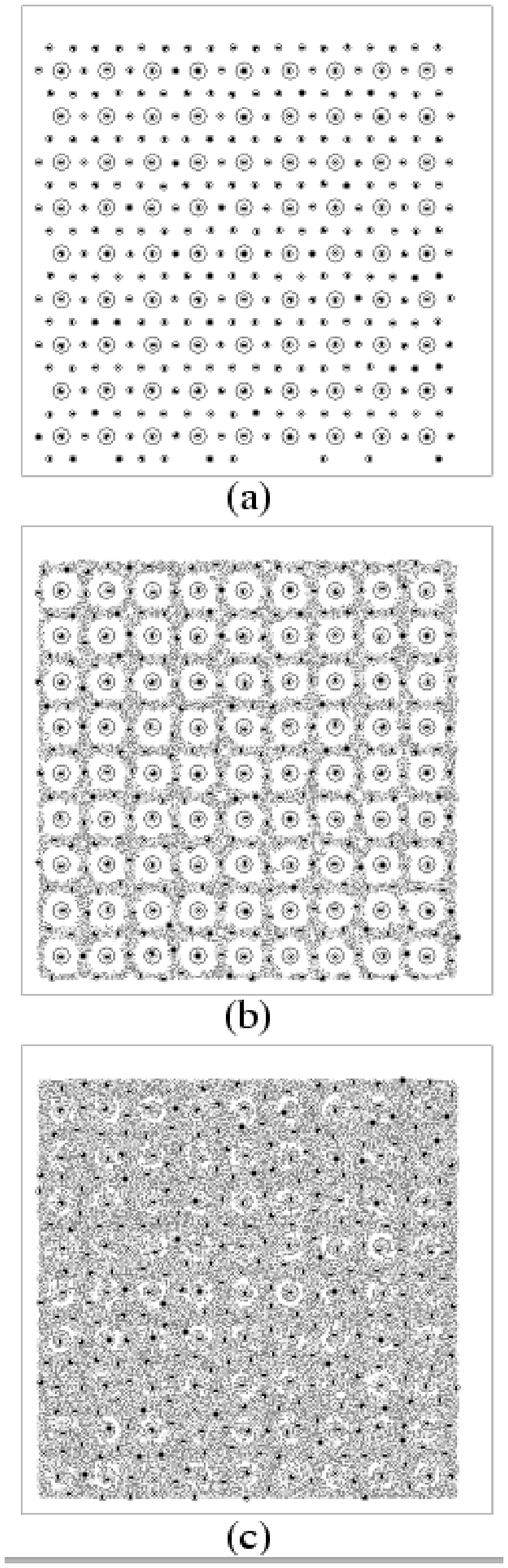}} 
\caption{
(a) The vortex positions (black circles) and pinning sites (open circles) for
$B/B_{\phi} = 4.0$ for a system with a square pinning array at 
$T = 0.001$. Here the vortex system is in a crystalline state. (b)
For $T=0.004$ the vortex trajectories show that the interstitial vortices 
are in a liquid state while the vortices at the pinning sites remain
immobile. (c) At $T = 0.01$ all the vortices are mobile.}  
\label{fig6}
\end{figure}

\begin{figure}
\centerline{
\epsfxsize=3.5in
\epsfbox{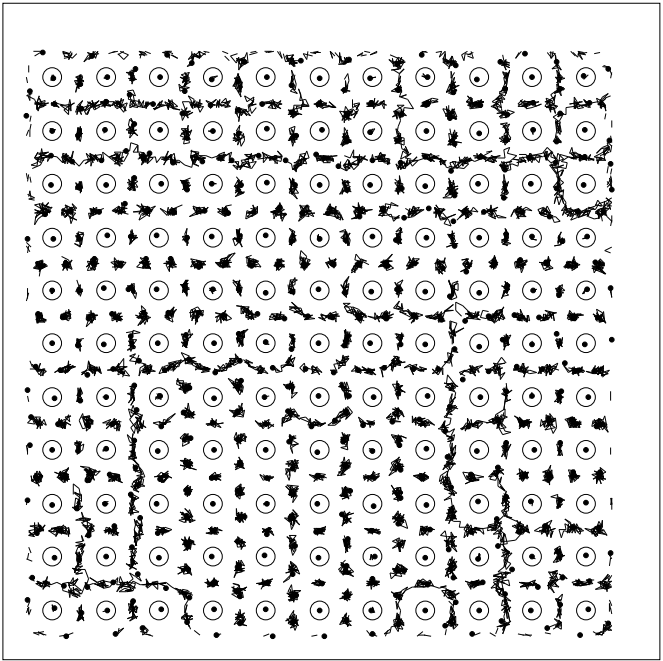}}
\caption{
The vortex positions (black circles) and pinning sites (open circles) for
a system with the same pinning parameters as in Fig.~6 for 
$B/B_{\phi} = 4.1$. Here $T = 0.001$ and  
the vortex trajectories show a portion of the 
interstitial vortices are mobile.}
\label{fig7}
\end{figure}  

\begin{figure}
\centerline{
\epsfxsize=3.5in
\epsfbox{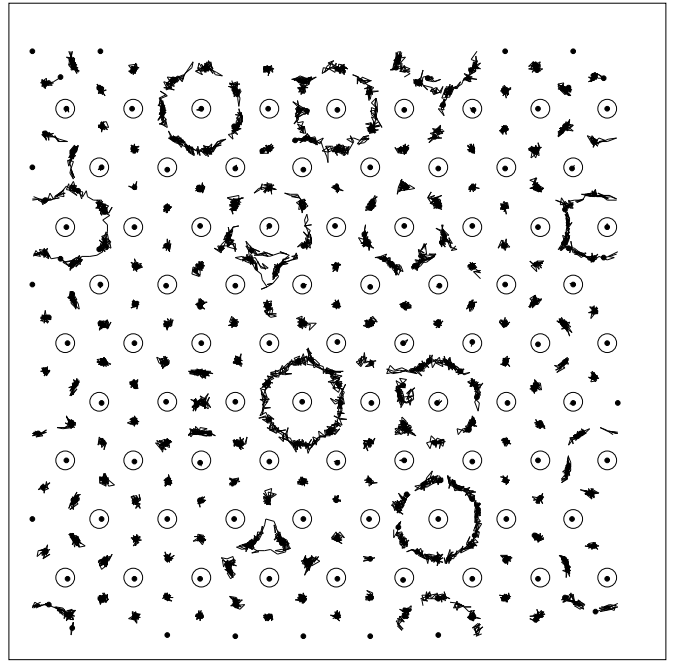}}
\caption{
The vortex positions (black circles) and pinning sites (open circles) for
a system with same parameters as in Fig.~6 with a triangular pinning array
with $B/B_{\phi} = 3.08$ and $T = 0.001$. Here the vortex motion can be
seen to occur in a ring excitation where vortices can rotate around 
pinned vortices. The ring excitations occur where there are seven vortices
around a pinned vortex. A second type of excitation can be seen in the 
trianglular trajectories where three vortices can rotate.}
\label{fig8}
\end{figure}

\end{document}